\documentclass[11pt]{article}
\usepackage{booktabs}

\usepackage[final]{acl}

\usepackage{times}
\usepackage{latexsym}
\usepackage{amsmath}
\usepackage{todonotes}

\usepackage[T1]{fontenc}

\usepackage[utf8]{inputenc}

\usepackage{microtype}
\usepackage{algorithm}
\usepackage{algorithmicx}
\usepackage{algpseudocode}

\usepackage{inconsolata}

\usepackage{graphicx}

\usepackage{listings}
\usepackage[table]{xcolor}
\definecolor{lightblue}{RGB}{235,242,249}
\definecolor{lightred}{RGB}{250,230,235}
\definecolor{jpblue}{RGB}{33, 99, 154}
\definecolor{jpred}{RGB}{188, 0, 45}

\title{\textsc{FinCARDS}: Card-Based Analyst Reranking\\ for Financial Document Question Answering}

\author{
\textbf{Yixi Zhou}\textsuperscript{1}\thanks{Equal contribution} \
\textbf{Fan Zhang}\textsuperscript{2}\footnotemark[1]\
\textbf{Yu Chen}\textsuperscript{2}\thanks{Corresponding author} \
\textbf{Haipeng Zhang}\textsuperscript{1}\footnotemark[2] \\
\textbf{Preslav Nakov}\textsuperscript{3} \
\textbf{Zhuohan Xie}\textsuperscript{3} \\
\textsuperscript{1}ShanghaiTech University \
\textsuperscript{2}The University of Tokyo \
\textsuperscript{3}MBZUAI\\
\{zhouyx2022, zhanghp\}@shanghaitech.edu.cn\\
\{zhang-fan@g.ecc, chen@edu.k\}.u-tokyo.ac.jp\\
\{preslav.nakov, zhuohan.xie\}@mbzuai.ac.ae}

\begin{document}
\maketitle

\begin{abstract}
Financial question answering (QA) over long corporate filings requires evidence to satisfy strict constraints on entities, financial metrics, fiscal periods, and numeric values. However, existing LLM-based rerankers primarily optimize semantic relevance, leading to unstable rankings and opaque decisions on long documents. We propose \textsc{FinCards}, a structured reranking framework that reframes financial evidence selection as \emph{constraint satisfaction} under a finance-aware schema.
\textsc{FinCards} represents filing chunks and questions using aligned schema fields (entities, metrics, periods, and numeric spans), enabling deterministic field-level matching. Evidence is selected via a multi-stage tournament reranking with stability-aware aggregation, producing auditable decision traces. Across two corporate filing QA benchmarks, \textsc{FinCards} substantially improves early-rank retrieval over both lexical and LLM-based reranking baselines, while reducing ranking variance, without requiring model fine-tuning or unpredictable inference budgets.
Our code is available at \url{https://github.com/XanderZhou2022/FINCARDS}.
\end{abstract}

\maketitle

\section{Introduction}

\begin{figure}[h]
  \centering
  \includegraphics[width=\linewidth]{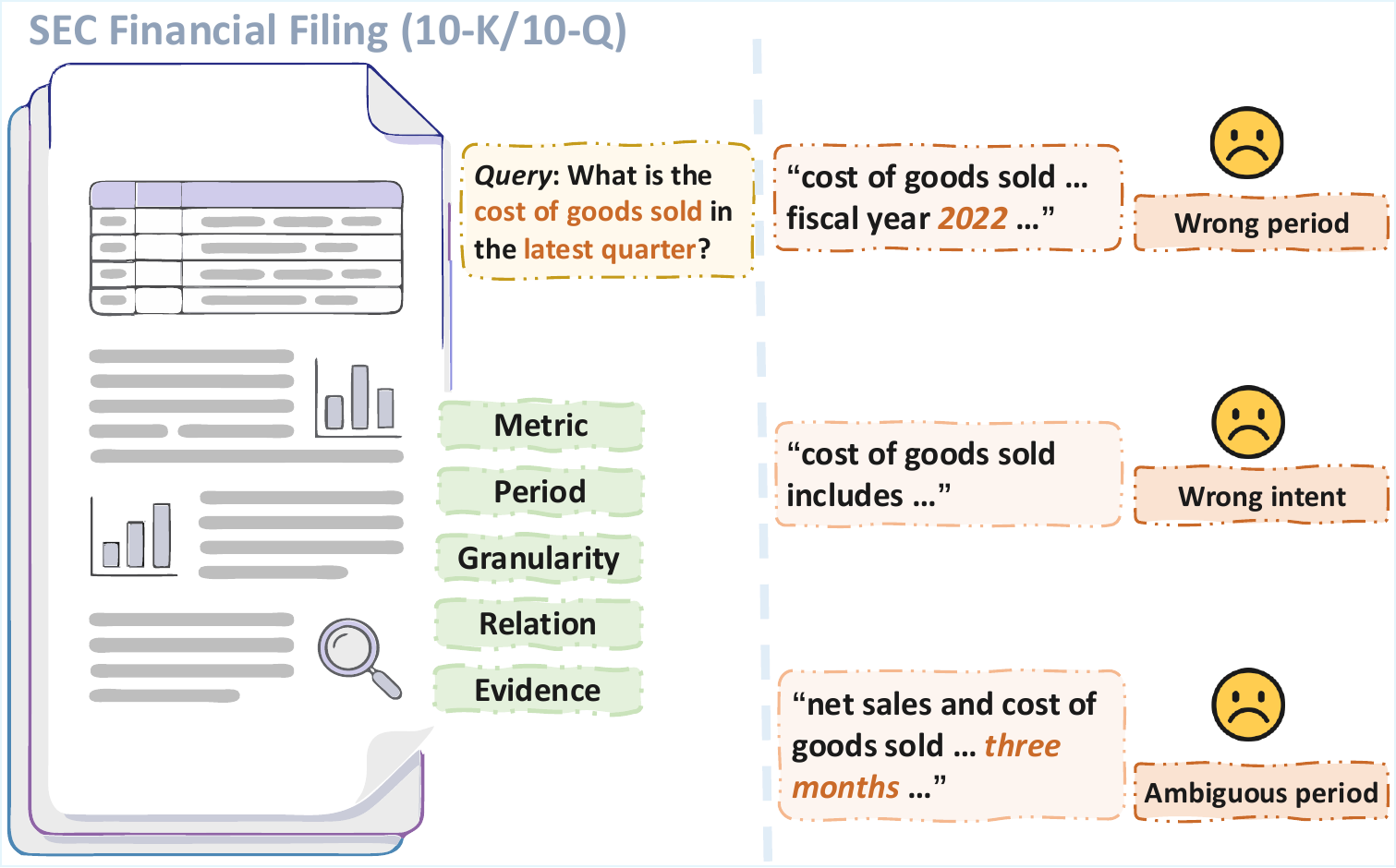}
  \caption{\textbf{Key challenge in financial QA.}
  Reranking must satisfy the correct metric and fiscal period (often numeric), not just semantic relevance. The illustration is based on U.S. SEC corporate financial filings and shows typical failure modes.}

  \label{fig:taskdef}
\end{figure}

Financial question answering (QA) over corporate filings is often framed as a retrieval problem, but in practice it is reranking under strict financial constraints. Correct evidence must simultaneously match the queried metric, fiscal period, and entity, and often must include an explicit numerical value.
These signals are sparsely distributed within hundreds of pages and are heavily interleaved with boilerplate disclosures and recurring statements. As a result, the problem reduces to reliably reranking candidate passages within a single document (Figure~\ref{fig:taskdef}) so that the top results satisfy all conditions~\citep{finqa,convfinqa,tatqa}.

A common agent-style approach is to feed large batches of text chunks into a large
language model (LLM) and ask it to rank or select evidence.
In long corporate filings, this approach breaks down for two practical reasons. First, \emph{scale}: multi-hundred-page reports quickly exceed LLM context budgets,
and expanding the input window leads to prohibitive token and latency costs. Second, \emph{opacity}: monolithic, prompt-driven rankings provide little insight into why a particular passage is selected, making the decisions difficult to inspect or audit, which is unacceptable in regulated financial analysis.
This leads to systematic errors such as selecting evidence from the wrong fiscal period, misaligning the metric of the query, or returning temporally ambiguous passages~\citep{sun2023rankgpt,ma2023lrl,choi2025finagentbench}.
As a result, generic LLM rerankers tend to optimize surface-level semantic relevance rather than explicit constraint satisfaction, making their decisions brittle in financial settings.

Unlike generic LLM rerankers that focus on semantic relevance, \textsc{FinCARDS} aligns evidence using explicit financial fields such as metrics and periods, producing auditable decision traces.

Our design is motivated by how financial analysts reason over long filings: evidence is not evaluated all at once, but progressively narrowed, ordered, and adjudicated under explicit criteria.
A junior analyst first screens likely evidence, a senior analyst makes a coherent global ordering, and a committee resolves close calls when distinctions are subtle.
We operationalize this workflow as a machine-usable alignment contract.
Document chunks are abstracted into structured \emph{cards} under a shared schema, questions are mapped to explicit intent specifications over the same fields, and a lightweight tournament-style review performs screening, global ordering, and adjudication to produce a ranked set of evidence passages.
This staged formulation improves numeric and temporal grounding, supports auditability, and keeps computation within predictable cost budgets.

Our focus is the \emph{intra-document} ranking setting: given a single, pre-selected filing, surface the most relevant chunks for a question. This isolates the core retrieval bottleneck in financial QA: locating grounded evidence within long documents before generation \citep{tatqa,finqa}.

We make the following contributions:
\begin{itemize}
    \item We reformulate financial question answering over long corporate filings as an \emph{intra-document evidence reranking} problem under strict numeric, temporal, and entity constraints, shifting the modeling focus away from monolithic long-context reasoning.
    \item We propose \textsc{FinCARDS}, a structured representation that abstracts document chunks into auditable evidence cards and maps questions into explicit intent specifications, enabling deterministic and interpretable alignment.
    \item We introduce a zero-shot, tournament-style reranking pipeline that produces stable ranked evidence sets under finance-aware criteria, and demonstrate consistent improvements in early precision over strong baselines.
\end{itemize}

\section{Related Work}
\paragraph{Financial QA benchmarks and evidence structure.}
Financial QA benchmarks have progressively shifted toward explicit modeling of evidence structure. Early work emphasized classification and extraction tasks, such as sentiment analysis and entity recognition, where evidence was implicit and locally contained \citep{araci2019finbert}.


Later benchmarks focused on structured, multi-step reasoning. FinQA \citep{finqa} and TAT-QA \citep{tatqa} formalized numerical and hybrid text-table reasoning, while ConvFinQA \citep{convfinqa} and FinChain \citep{finchain} extended this to conversational settings and verification of intermediate reasoning steps. Recent suites such as FinBen \citep{finben} and PIXIU \citep{pixiu} consolidated these into evidence-sensitive multi-task benchmarks, with multilingual extensions including CFinBench \citep{cfinbench} and Plutus \citep{plutus}. Despite this progress, retrieval and evidence selection remain challenging, as shown by FinanceBench \citep{islam2023financebench}, FinDER \citep{choi2025finder}, and FinAgentBench \citep{choi2025finagentbench}, motivating our focus on intra-document evidence structure.


\paragraph{Retrieval and LLM reranking.}
Classical IR relies on lexical methods such as BM25, while modern pipelines incorporate dense retrievers and cross-encoders \citep{xiong2021ann}. Recently, LLMs have been used as zero-shot rerankers: listwise approaches like RankGPT and LRL show instruction-tuned LLMs can reorder candidates competitively without task-specific training \citep{sun2023rankgpt,ma2023lrl, li2026retrieval}. However, listwise prompts can be input-order sensitive and context-length constrained. Pairwise prompting (\emph{A vs.\ B?}) improves calibration and stability \citep{qin2024prp}, while setwise/tournament strategies mitigate order sensitivity and scale better with long lists \citep{zhuang2023setwise,chen2024tourrank}. Simple rank fusion such as reciprocal rank fusion (RRF) remains a strong baseline to aggregate noisy rankings \citep{cormack2009rrf}. Recent work suggests listwise and pairwise comparisons play complementary roles in robust reranking, from global ordering to resolving close decisions. Our work builds on these insights and adapts them to the financial domain by structuring comparisons within a constraint-driven review process that enforces explicit numeric and temporal alignment \citep{xing2025llmrankreliability}.

\paragraph{Agentic reasoning and deliberative inference.}
A line of work studies how large language models emulate multi-step reasoning and decision-making processes. 
ReAct interleaves reasoning with external actions \citep{yao2022react}, Tree-of-Thoughts explores multiple solution paths through structured search \citep{yao2023tot}, and Reflexion introduces self-critique with episodic memory for iterative improvement \citep{shinn2023reflexion}. 

Self-consistency further improves reasoning reliability by aggregating diverse reasoning paths \citep{wang2023selfconsistency, liu2026beyondinvestor}. 
These approaches provide general mechanisms for structured reasoning, but they primarily focus on generation and problem solving rather than evidence selection under domain-specific constraints.

 \paragraph{Model-based evaluation and RAG-based systems.}
Another line of work examines reliability in model-based evaluation, highlighting
variance and calibration challenges in LLM-as-a-judge settings.
In document-centric tasks, layout-aware models such as LayoutLMv3 improve
understanding of visually structured financial documents \citep{huang2022layoutlmv3},
while benchmarks such as ChartQA and TAT-QA emphasize hybrid reasoning over
textual and tabular content \citep{masry2022chartqa,tatqa}.
Related efforts in structured or multi-agent reasoning systems further explore how
models can coordinate multiple reasoning steps or agents for complex tasks
\citep{xu2026rcbsfmultiagent, zhou2026sqlstructeval}.

Surveys of retrieval-augmented generation (RAG) show that system performance is
often limited by retrieval quality rather than model capacity \citep{fan2024rag}.
Building on this observation, recent work studies retrieval-aware reasoning and
information planning, where models dynamically decide what to retrieve and how
to use retrieved evidence \citep{li2026modelun}. Other approaches investigate
dynamic or uncertainty-aware retrieval strategies that adapt retrieval behavior
to evolving contexts \citep{xu2026selfcorrectingrag}. In parallel, multi-agent
frameworks explore how multiple specialized components can coordinate to handle
complex reasoning tasks \citep{yang2026mat, ma2026medla}, while domain-specific
applications further demonstrate the effectiveness of structured, information-driven
reasoning pipelines \citep{Pei2025vcp, Liu2026fine}.


In this context, our work draws inspiration from these directions but focuses specifically on improving intra-document retrieval reliability and robustness in financial QA settings. Rather than expanding model capacity or relying on implicit reasoning, we instead introduce structured intermediate representations and a carefully designed staged reranking process that enforces explicit numeric and temporal constraints, thereby directly addressing the alignment and stability challenges identified in prior RAG and model-based evaluation studies.

\begin{figure*}[t]
  \centering
  \includegraphics[width=\textwidth]{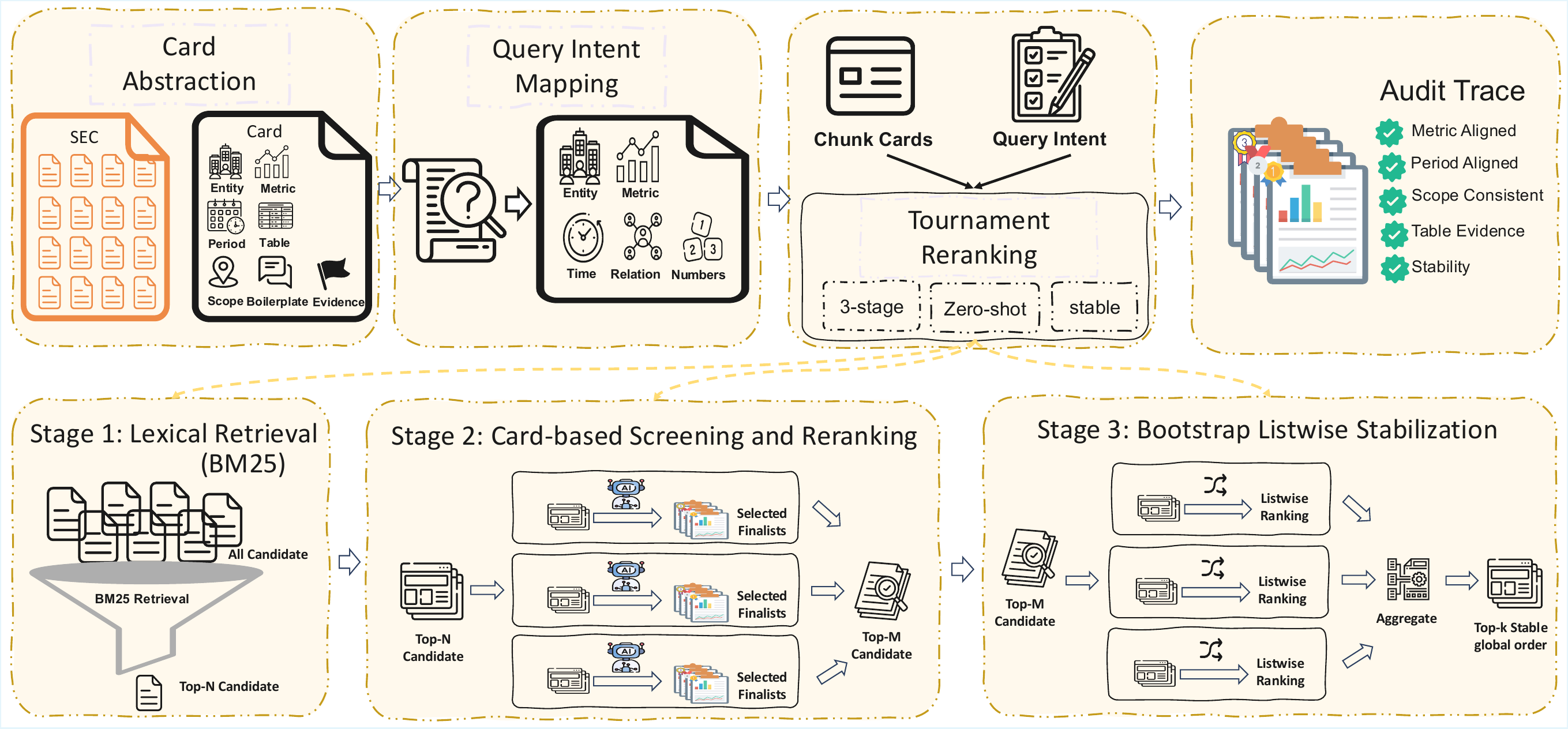}
  \caption{\textbf{Overview of the \textsc{FinCARDS} pipeline.}
From a SEC filing and a user question, the system constructs
structured Cards, a structured query intent, and a tournament reranking module
that produces the final Top-$k$ evidence chunks.}
  \label{fig:pipeline}
\end{figure*}

\section{\textsc{FinCARDS}}
\label{sec:method}


In this section, we first provide an overview of our method. We then describe each component of the framework, including the representations and overall reranking procedure.

\subsection{Overview}
\label{sec:overview}

We study \emph{intra-document evidence reranking} for financial question answering: given a user question and all chunks from a single long SEC filing (e.g., 10-K/10-Q), the goal is to rank chunks so that the top-$k$ results satisfy the required financial conditions: metric, fiscal period, entity, and often explicit numbers.

Figure~\ref{fig:pipeline} summarizes our approach, \textsc{FinCARDS}, which decomposes this problem into three components. \textbf{(1)~Card abstraction} converts each chunk into a compact, structured \emph{card} that records finance-relevant fields (entities, metrics, periods, numbers, and section cues) for auditable matching. \textbf{(2)~ Query intent mapping} converts the question into a structured intent that specifies the demanded entities/metrics/periods and whether numeric evidence is required. \textbf{(3)~Tournament reranking} performs staged, zero-shot reranking over cards, combining a screening step, a global listwise ordering step, and a targeted adjudication step, followed by lightweight fusion and post-hoc alignment to produce the final top-$k$ evidence list.

In addition to the final ranking, the pipeline produces an explicit \emph{audit trace} for each selected chunk, recording which Card fields were matched, how the chunk was retained or filtered at each stage, and how its final rank was determined.

\subsection{Card Abstraction}
\label{sec:card}

\paragraph{Why Card Abstraction.}
The \emph{card abstraction} module makes financial evidence explicit and auditable
before any ranking decisions are made.
It converts raw text chunks from long SEC filings into compact, structured
records that explicitly encode entities, financial metrics, fiscal periods,
and verbatim numeric spans.
In the following stages, we
compare candidates through field-level matching instead of free-form
semantic similarity.

By operating on fields instead of raw text, downstream stages can perform reliable comparisons under strict numeric and temporal constraints.
The corresponding Card instantiation prompt is provided in Appendix~\ref{app:cardprompts}.

\paragraph{Motivation.}
This design directly addresses recurring failure modes in financial QA over long
filings.
Lexical retrieval is particularly vulnerable to numeric drift, temporal
misalignment, and boilerplate repetition, where legally mandated disclosures
dominate surface signals without conveying substantive evidence.
Card abstraction mitigates these issues by enforcing explicit temporal
normalization, verbatim numeric copying, and boilerplate awareness at the
representation level.

\paragraph{Analyst analogy.}
Conceptually, card abstraction mirrors the preparatory work of a junior financial
analyst: relevant numbers, periods, and contextual cues are recorded explicitly
so that subsequent reviewers can assess relevance without repeatedly
reinterpreting raw text.
Crucially, this transformation also changes the information flow in downstream
reasoning. Instead of operating over raw, heterogeneous text, later stages
interact with a normalized representation space in which key financial attributes
are explicitly surfaced and aligned. This reduces ambiguity in comparison and
enables consistent handling of semantically equivalent but lexically diverse
expressions (e.g., abbreviations or paraphrased financial terms).
The resulting Card corpus therefore defines a structured evidence space that
supports reliable field-level alignment in the tournament-style reranking stages
described in Section~\ref{sec:reranking}.

\paragraph{Formal definition.}
Let $\mathcal{X}=\{x_1,\dots,x_L\}$ denote the text chunks from a single SEC filing.
Each chunk $x_i$ is mapped to a structured \emph{Chunk Card} via a
schema-constrained extraction function with deterministic decoding:
\begin{equation}
f:\mathcal{X}\!\to\!\mathcal{C},\qquad
c_i=f(x_i)=\mathrm{ChunkCard}(x_i).
\end{equation}

Each Chunk Card $c_i\in\mathcal{C}$ is a typed record with a full schema, from
which we derive a compact \emph{alignment core} for intent matching, and the
remaining auxiliary fields are used only for screening and stability control:
\begin{equation}
\begin{aligned}
c_i &= \big(c_i^{\text{core}}, c_i^{\text{aux}}\big), \\
c_i^{\text{core}} &= (T_i, E_i, M_i, N_i, P_i, S_i, D_i, \Xi_i).
\end{aligned}
\end{equation}

The core schema $c_i^{\text{core}}$ is an \emph{alignment core (projection)} of the
full Chunk Card, used exclusively for intent alignment.
Auxiliary fields $c_i^{\text{aux}}$ are never used for alignment and are used
only for screening and stability control.

\paragraph{Core fields.}
$T_i\!\in\!\mathcal{T}$ is a topic label,
$E_i\!\subseteq\!\mathcal{E}$ is the set of explicitly mentioned entities,
$M_i\!\subseteq\!\mathcal{M}$ are financial metrics,
$N_i$ is a (possibly empty) multiset of verbatim numeric spans,
$P_i\!\in\!\mathcal{P}$ is a normalized fiscal period or interval,
$S_i\!\in\!\mathcal{S}$ is the section identifier,
$D_i\!\in\!\mathcal{D}$ is a derived entity--metric--period triple,
and $\Xi_i$ are evidence spans linking all fields to the original text
(\emph{audit trace}).

\paragraph{Auxiliary fields.}

$c_i^{\text{aux}}$ includes cues such as scope descriptors, table signatures, and a boilerplate flag. These are derived via rules over $x_i$ and guide Stage~2 and~3 screening and stability control.

\subsection{Query Intent Mapping}
\label{sec:query}

On the query side, we map each natural language question to a structured
\emph{intent} representation that explicitly encodes entities, metrics, temporal
constraints, and numeric requirements.
This representation enables direct field-level matching against Card schemas,
rather than relying on unconstrained text similarity.

Financial questions are often underspecified in surface form.
For example, the query ``\emph{How did revenue change last quarter?}'' implicitly
requires numeric evidence, a specific fiscal period, and a comparison relation.
Intent mapping resolves this ambiguity by decomposing questions into dimensions
that can be directly aligned with Card fields.

Each question $q_j$ is mapped to an intent object:
\begin{equation}
\mathrm{Intent}(q_j) =
\big(T_j, E_j, M_j, R_j, \Theta_j, \nu_j, K_j\big),
\end{equation}
where $T_j$ is a topic label, $E_j$ entities, $M_j$ metrics,
$R_j$ the relation type (e.g., comparison or trend),
$\Theta_j$ temporal constraints,
$\nu_j$ whether explicit numeric evidence is required,
and $K_j$ lexical keywords.

Together, Card abstraction and intent mapping expose a shared schema that
enables explicit field-level alignment between query requirements and candidate
evidence in the tournament reranking stages (Section~\ref{sec:reranking}).
By operating over aligned representations on both the query and document sides,
the framework reduces reliance on unconstrained semantic matching and instead
grounds decisions in structured financial attributes such as metrics, periods,
and scope.
The corresponding query intent prompt is provided in Appendix~\ref{app:queryprompts}.

\subsection{Tournament Reranking}
\label{sec:reranking}

We now describe the tournament-style reranking module used to select evidence
within a single filing.
The pipeline assumes (\emph{i})~a Card corpus $\mathcal{C}$ derived from filing chunks
(Section~\ref{sec:card}), and (\emph{ii})~a structured query intent
$\mathrm{Intent}(q)$ extracted from the question (Section~\ref{sec:query}).

Ranking proceeds in three stages: recall-oriented candidate generation,
Card-based semantic filtering, and stability-aware listwise aggregation.
This staged design reflects how financial analysts progressively narrow, order,
and adjudicate evidence under strict numeric and temporal constraints.

\subsubsection{Preliminaries}

Given a question $q$ and a long-text SEC filing, let
$\mathcal{X}=\{x_1,\dots,x_L\}$ denote the set of text chunks and let
$\mathcal{C}=\{c_1,\dots,c_L\}$ be the corresponding Chunk Cards.
We extract a structured intent representation $\mathrm{Intent}(q)$ as described
in Section~\ref{sec:query}.

The goal of tournament reranking is to return an ordered list of chunk indices
$\pi=(\pi_1,\dots,\pi_k)$ corresponding to the top-$k$ evidence chunks in the
filing, optimized for early-rank relevance.

\subsubsection{Stage~1: Lexical Retrieval (BM25)}

Stage~1 constructs a high-recall candidate set via lexical retrieval within the same filing. Concretely, we score all chunks using the BM25 ranking function \citep{robertson2009probabilistic} and keep the top-$N$ candidates:
\begin{equation}
\mathcal{S}_1(q) = \mathrm{TopN}\big(\mathrm{BM25}(q, x_i)\big).
\end{equation}
To make the budget comparable across filings of different lengths, we use a length-adaptive cutoff
\begin{equation}
N = \mathrm{clamp}\!\left(\left\lceil rL \right\rceil,\; N_{\min},\; N_{\max}\right),
\end{equation}
with $r=0.5$, $N_{\min}=60$, and $N_{\max}=150$ (and $N=L$ if $L<N_{\min}$).
Stage~1 serves as a recall-oriented starting point for downstream reranking and
highlights cases where relevant evidence receives low lexical scores.

\subsubsection{Stage~2: Card-Based Screening and Reranking}

Stage~2 reduces the Stage~1 candidate set using structured Card representations,
without accessing raw chunk text.
Starting from $\mathcal{S}_1(q)$, we partition the candidates into groups via
round-robin assignment so that each individual group contains a mix of high-, mid-, and
low-ranked Stage~1 candidates.
Let $\{\mathcal{G}_1,\dots,\mathcal{G}_m\}$ denote groups of size approximately $g$
(default $g=25$).

For each group $\mathcal{G}_t$, an LLM agent selects a small set of finalists:
\begin{equation} \mathcal{F}_t = \textsc{Select}(\mathcal{G}_t, q), \; |\mathcal{F}_t| \in [k_{\min}, k_{\max}]. \end{equation}

The selection rubric emphasizes alignment between the query intent and Card fields,
including
(\emph{i})~metric overlap,
(\emph{ii})~temporal compatibility,
(\emph{iii})~scope consistency (company-wide vs.\ segment/region/product),
(\emph{iv})~appropriate evidence type (e.g., table-centric cards for quantitative queries),
and
(\emph{v})~down-weighting boilerplate content unless it is uniquely relevant.

To avoid over-filtering, we enforce lightweight coverage constraints.
For instance, trend queries must retain at least one temporally grounded,
table-bearing card, while definition or policy queries must retain at least one
explanatory card.

We merge group-level selections by union and deduplication,
retaining the highest relevance score when a candidate appears multiple times:
\begin{equation}
\mathcal{S}_2(q) = \mathrm{Dedup}\!\left(\bigcup_{t=1}^{m}\mathcal{F}_t\right).
\end{equation}
The Stage~2 batch selection prompt is provided in Appendix~\ref{app:stage2_batch_prompt}.

\subsubsection{Stage~3: Bootstrap Listwise Stabilization}


While Stage~2 reduces the candidate set using structured Card cues, the resulting ranking can be unstable due to sensitivity to single-pass LLM judgments to grouping and comparison context. Stage~3 addresses this by enforcing \emph{ranking stability} through multi-round bootstrap listwise aggregation.


Given the Stage~2 candidate set $\mathcal{S}2(q)$ of size $M$, we select a group size $g\in[15,25]$ (adapted to $M$) and perform up to $R{\max}=5$ bootstrap rounds. In each round $r$, candidates are shuffled with different seeds and partitioned into groups ${\mathcal{H}{r,1},\dots,\mathcal{H}{r,p_r}}$. For each group, an LLM produces a listwise ranking. Random regrouping exposes each candidate to multiple comparison contexts, mitigating bias from any single partition.


To aggregate rankings across groups and rounds, we use \emph{normalized Borda scores}. This choice is motivated by two considerations: (\emph{i})~Borda aggregation preserves fine-grained relative ordering information rather than relying on hard selection or voting, and (\emph{ii})~normalization ensures comparability across groups of different sizes that arise under bootstrap partitioning. For a group $\mathcal{H}$ of size $|\mathcal{H}|$, an item ranked at position $\rho$ receives a score
\begin{equation}
s(\rho;\mathcal{H})=\frac{|\mathcal{H}|-\rho}{|\mathcal{H}|-1}\in[0,1].
\end{equation}

We then accumulate the scores over all appearances of each candidate:
\begin{equation}
S(i\mid q)=\sum_{r}\sum_{t} s\!\left(\rho_{r,t}(i);\mathcal{H}_{r,t}\right).
\end{equation}
Sorting by $S(i\mid q)$ yields a stable global ranking $\pi$,
from which we obtain the top-$k$ evidence chunks.
The Stage~3 listwise ranking prompt is provided in Appendix~\ref{app:stage3_listwise_prompt}.

To control the computational cost, we monitor the convergence of the current top-$k$ set. If the Jaccard similarity between the top-$k$ results from consecutive rounds exceeds a threshold (0.9), the procedure terminates early. Early stopping is based solely on the consistency of predicted rankings, without access to any ground-truth labels.

\subsubsection{Reproducibility and Cost}

All LLM interactions use deterministic decoding and strict
JSON schema validation to ensure reproducibility.
Prompt templates and structured interfaces are detailed in Appendix~\ref{app:prompts}.
In addition to the final Top-$k$ ranking, the pipeline outputs an explicit
\emph{audit trace} for each selected chunk, recording
stage-wise candidate lists, grouping decisions, and per-round ranks in Stage~3.
This trace exposes which structured Card fields (e.g., metric, period, scope,
and table cues) were matched, how each chunk was retained or filtered across
stages, and how its final rank was determined, enabling transparent inspection
and controlled multi-model comparisons under identical algorithms and prompts.

In terms of cost, let $L$ be the number of chunks in a filing,
$N$ be the Stage~1 candidate cutoff, and $M = |\mathcal{S}_2(q)|$ be the Stage~2 candidate size.
Stage~1 scores all chunks using BM25 in $O(L)$ time per query.
Stage~2 requires $m = \lceil N / g \rceil$ LLM calls (one per group),
and Stage~3 performs at most $R_{\max}$ bootstrap rounds with
$\lceil M / g \rceil$ listwise calls per round.
In practice, the total number of LLM calls is often substantially reduced
by early stopping once the Top-$k$ ranking stabilizes.
Detailed token-level cost analysis is provided in Appendix~\ref{app:token_cost}.

The three-stage pipeline implements a refinement
process that decomposes evidence selection into distinct and complementary
subproblems. Stage~1 retrieves a high-recall candidate set under lexical signals,
Stage~2 enforces structured semantic alignment through explicit constraint matching,
and Stage~3 resolves residual uncertainty via repeated comparison and aggregation.

This design matters in long financial documents, where a single-stage approach
either misses relevant evidence due to limited recall or produces unstable rankings
over large candidate sets. By separating recall, alignment, and stabilization, the
pipeline assigns each stage a well-defined role and avoids overloading any single
decision step, which leads to both higher accuracy and more reliable ranking behavior.

\section{Experiments and Evaluation}

This section presents the experimental evaluation of our proposed framework. We first describe the experimental setup, followed by the different system variants used for comparison and the evaluation measures. We then report the main results and offer a detailed performance analysis to understand the contribution of each component.

\subsection{Experimental Setup}

We evaluate our multi-stage intra-document retrieval and ranking system on the
\textbf{FinAgentBench} \cite{choi2025finagentbench} benchmark, which consists of financial question answering tasks
derived from U.S.\ SEC filings (10-K and 10-Q).
For each query, the system is provided with a \emph{single financial document}
and must identify and rank the most relevant evidence chunks within that document.
This setting isolates the challenge of \emph{intra-document retrieval}, where
relevant evidence is often sparse, temporally constrained, and interleaved with
boilerplate disclosures.

All experiments follow a consistent and controlled inference protocol with deterministic decoding
and structured outputs, ensuring reproducibility and fair comparison across all system variants.

\subsection{System Variants}
We compare traditional lexical retrieval, zero-shot LLM-based reranking,
and several variants of our multi-stage pipeline.
All systems share the same document chunking and evaluation protocol.
\textbf{Stage~1} refers to BM25-based lexical retrieval. For fairness, the zero-shot LLM reranking baseline operates on the same Stage~1 candidate set, with identical candidate budgets. All reranking-based methods (including our pipeline and cross-encoder baselines) operate on the same Stage~1 candidate pool to ensure fair comparison.
The controlled candidate-pool protocol is further detailed in Appendix~\ref{app:candidate_fairness}, and additional stronger baselines are reported in Appendix~\ref{app:stronger_baselines}.

\paragraph{Zero-shot LLM Reranking.}
A standard LLM-based reranking baseline that applies a single-pass,
zero-shot LLM to reorder the \emph{Stage~1 candidate set} using raw chunk text.
Unlike our approach, this baseline does not leverage structured Card representations, candidate grouping, or multi-round stabilization.

\paragraph{Stage~1 + Stage~3.}
An ablated variant that applies the Stage~3 bootstrap listwise ranking
directly on the Stage~1 candidate set, without Card-based filtering.
This setting isolates the effect of stability-aware ranking independent
of structured semantic screening.

\paragraph{Stage~1 + Stage~2.}
A two-stage variant that applies Card-based semantic screening and reranking
on top of Stage~1 retrieval, but does not include the bootstrap-based
stability mechanism of Stage~3.

\paragraph{Stage~1 + Stage~2 + Stage~3 (Full Pipeline).}
Our full system, which sequentially combines Stage~1 lexical retrieval,
Stage~2 Card-based semantic filtering, and Stage~3 bootstrap listwise
stabilization.

\subsection{Evaluation Measures}

We adopt standard information retrieval metrics at rank~10, as each question in FinAgentBench is typically associated with a small set of relevant evidence chunks (on the order of ten), making early-rank quality the primary evaluation focus in our setting. \textbf{nDCG@10} measures graded relevance with emphasis on early ranks, \textbf{MAP@10} captures precision across the top-ranked results, and \textbf{MRR@10} reflects how quickly the first relevant evidence chunk appears. All measures are reported as averages over all evaluation queries, with scores multiplied by 100 and reported as percentages for consistency.

\subsection{Main Results}

Table~\ref{tab:mainresults} summarizes the main results on FinAgentBench
under the intra-document retrieval setting. Our three-stage pipeline achieves a large and consistent improvement over both traditional and LLM-based baselines across all metrics.

Compared to Stage~1, the full pipeline improves nDCG@10 by over
27 points and MRR@10 by nearly 20 points, absolute.
Even relative to a strong zero-shot LLM reranking baseline,
our approach yields substantial gains (+15.8 nDCG@10),
demonstrating that na\"{i}ve LLM reranking is insufficient for financial evidence selection.

Importantly, these accuracy gains are achieved while
\emph{progressively reducing the candidate set size}
from roughly 100 chunks to fewer than 25,
indicating that the proposed design improves both ranking quality
and retrieval efficiency.

\begin{table}[t]
\centering
\scriptsize
\setlength{\tabcolsep}{2pt}
\renewcommand{\arraystretch}{1.05}
\begin{tabular}{@{}p{0.34\columnwidth}cccc@{}}
\toprule
System & nDCG@10 & MAP@10 & MRR@10 & Cand. Size \\
\midrule
\rowcolor{lightblue}
\multicolumn{5}{c}{\textit{Traditional Baseline}} \\
Stage~1 & 44.26 & 60.01 & 69.88 & 25 \\
\midrule
\rowcolor{lightblue}
\multicolumn{5}{c}{\textit{LLM-based Baseline}} \\
Zero-shot LLM Reranking & 55.80 & 66.50 & 78.20 & $100\!\to\!25$ \\
\midrule
\rowcolor{lightblue}
\multicolumn{5}{c}{\textit{Ablation of Our Method}} \\
Stage~1+Stage~3 & 58.23 & 68.42 & 77.56 & $100\!\to\!25$ \\
Stage~1+Stage~2 & 63.66 & 73.09 & 82.95 & $100\!\to\!40$ \\
\midrule
\rowcolor{lightblue}
\multicolumn{5}{c}{\textit{Our Full Pipeline}} \\
Stage~1+Stage~2+Stage~3 &
\textbf{71.58} & \textbf{78.69} & \textbf{89.17} & $100 \xrightarrow{40} 25$ \\
\bottomrule
\end{tabular}
\caption{\textbf{Main results on FinAgentBench under intra-document retrieval.} All retrieval metrics (nDCG@10, MAP@10, MRR@10) are reported as percentages.
The proposed three-stage pipeline substantially improves early-rank accuracy
while progressively reducing the candidate set size at each stage.}

\label{tab:mainresults}
\end{table}

\subsection{Analysis}






The results in Table~\ref{tab:mainresults} provide clear and consistent evidence that each stage of the proposed pipeline contributes meaningfully to retrieval effectiveness across all evaluation metrics and settings.

First, zero-shot LLM reranking improves over BM25, but remains limited, highlighting that replacing lexical scores with unstructured LLM judgments does not adequately resolve temporal mismatch, scope ambiguity, or boilerplate interference in financial filings, especially in long and noisy documents where surface similarity can be misleading.

Second, introducing Card-based filtering in Stage~2 yields a large performance jump (+7.9 nDCG@10 over zero-shot reranking), confirming that \emph{structured intermediate representations are crucial} for aligning query intent with financial evidence and enforcing explicit constraint satisfaction during selection in a more controlled and interpretable manner.

Finally, Stage~3 further improves early-rank metrics by stabilizing the rankings across multiple comparison contexts and reducing variance in model decisions across different runs. This demonstrates that \emph{ranking stability}, rather than additional semantic filtering alone, is essential for reliable early precision in long, noisy financial documents.

Overall, the analysis validates the core design principles of our proposed framework: structured reasoning, progressive filtering, and stability-aware aggregation, all achieved without task-specific fine-tuning or additional supervision, while remaining robust across different model backbones and evaluation conditions.

\subsection{Ablation Study}
\label{sec:ablation}

Below, we perform an ablation study in order to evaluate the impact of the individual components of our framework.

\subsubsection{Stage~2: Card Component Ablation}

We ablate the individual components of the Card representation used in Stage~2,
while keeping the pipeline and the model fixed. The results are shown in Table~\ref{tab:stage2_ablation}.

\begin{table}[t]
\centering
\small
\setlength{\tabcolsep}{3pt}
\renewcommand{\arraystretch}{1.05}
\begin{tabular}{lccc}
\toprule
Variant & nDCG@10 & MAP@10 & MRR@10 \\
\midrule
Full Card      & \textbf{63.66} & \textbf{73.09} & \textbf{82.95} \\
w/o temporal\_data        & 59.80 & 69.20 & 78.50 \\
w/o financial\_metrics    & 61.20 & 70.85 & 80.20 \\
w/o tables                & 58.50 & 67.80 & 77.20 \\
w/o scope                 & 62.15 & 71.80 & 81.50 \\
\midrule
Only summary              & 54.20 & 63.50 & 72.80 \\
Raw chunks (no Card)      & 49.50 & 60.80 & 70.15 \\
\bottomrule
\end{tabular}
\caption{\textbf{Stage~2 ablation results.} In the top part, each variant removes one component from the Card representation, while in the bottom part we show results when using just a summary or raw chunks.}
\label{tab:stage2_ablation}
\end{table}

We can see in Table~\ref{tab:stage2_ablation} that structured Card fields are essential for effective semantic filtering.
Removing temporal information causes the largest degradation, highlighting the importance of temporal alignment in financial QA.
Ablating table indicators or financial metrics also leads to substantial performance drops, indicating that identifying quantitative evidence is critical even without exposing raw numbers.
Using only free-text summaries performs poorly, and operating directly on raw chunks yields the worst results.

Overall, we can conclude that the Card abstraction is a necessary intermediate representation rather than a mere efficiency optimization.

\subsubsection{Stage~3: Ranking Strategy Ablation}

We further analyze Stage~3 by ablating key design choices in the bootstrap-based listwise ranking procedure.
In addition to ranking quality, we report \textit{rank variance} as a stability metric across bootstrap rounds (formalized in Appendix~\ref{app:rank_variance}). Specifically, for each candidate, we record its rank position in each round and compute the variance of these positions, then average over all candidates.

\begin{table}[t]
\centering
\scriptsize
\setlength{\tabcolsep}{3pt}
\renewcommand{\arraystretch}{1.05}
\begin{tabular}{@{}p{0.34\columnwidth}cccc@{}}
\toprule
Variant & nDCG@10 & MAP@10 & MRR@10 & Rank Var. \\
\midrule
Bootstrap (R=3--5)        & \textbf{71.58} & \textbf{78.69} & \textbf{89.17} & 0.0342 \\
Single Round (R=1)        & 68.20 & 75.20 & 85.80 & 0.0856 \\
Fixed Grouping            & 69.15 & 76.35 & 86.95 & 0.0621 \\
Mean Rank Aggregation     & 69.80 & 77.05 & 87.50 & 0.0498 \\
Voting Aggregation        & 67.35 & 74.80 & 85.10 & 0.0723 \\
No Early Stopping         & 71.65 & 78.75 & 89.25 & 0.0318 \\
\bottomrule
\end{tabular}
\caption{\textbf{Stage~3 ablation results.} We report early-rank retrieval quality and rank variance (lower is more stable), highlighting the role of bootstrap-based stabilization.}
\label{tab:stage3_ablation}
\end{table}

Table~\ref{tab:stage3_ablation} shows that bootstrap-based aggregation is crucial for both accuracy and stability.
Single-round ranking exhibits much higher variance, confirming the instability of one-shot LLM judgments.


Random regrouping outperforms fixed grouping, indicating diverse comparison contexts reduce bias. Disabling early stopping yields marginal gains while increasing computation, as rankings typically converge within 3--4 rounds. Overall, Stage~3 acts as a stability layer that improves early precision while reducing ranking variance.

\subsection{Robustness Across LLM Backbones}

To test model dependence, we run Stage~2 and Stage~3 with six LLM backbones. All other components of the pipeline remain unchanged.

\paragraph{Analysis.}
Table~\ref{tab:multi_model} shows that the proposed pipeline is robust across
diverse LLM backbones.

First, all models exhibit a substantial improvement from Stage~1 to Stage~2, indicating that Card-based semantic filtering consistently improves
evidence selection regardless of model capacity.
This suggests that the gains are primarily driven by the structured intermediate representation rather than backbone-specific reasoning ability.

Second, Stage~3 further improves or stabilizes ranking quality for every model,
with consistent gains in nDCG@10 and MRR@10.
This confirms that bootstrap-based listwise aggregation effectively mitigates
the variance of single-pass LLM rankings across architectures.

\begin{table}[t]
\centering
\small
\setlength{\tabcolsep}{4pt}
\renewcommand{\arraystretch}{1.05}
\begin{tabular}{@{}p{0.38\columnwidth}ccc@{}}
\toprule
Stage / Model & nDCG@10 & MAP@10 & MRR@10 \\
\midrule
\rowcolor{lightblue}
Stage~1 (BM25) & 44.26 & 60.01 & 69.88 \\
\midrule
\rowcolor{lightblue}
\multicolumn{4}{c}{\textit{Stage~2: Card-based Filtering}} \\
GPT-5 Mini              & 63.66 & 73.09 & 82.95 \\
GPT-4 Mini              & 64.63 & 75.63 & 84.66 \\
Claude-4.5-Opus           & \textbf{70.89} & \textbf{81.03} & \textbf{89.67} \\
Claude-4.5-Sonnet           & 67.15 & 77.16 & 85.71 \\
Gemini 2.5 Flash        & 63.28 & 73.32 & 82.46 \\
Gemini 3 Pro (Preview)  & 67.06 & 76.53 & 85.98 \\
\midrule
\rowcolor{lightblue}
\multicolumn{4}{c}{\textit{Stage~3: Bootstrap Stable Ranking}} \\
GPT-5 Mini              & 71.58 & 78.69 & 89.17 \\
GPT-4 Mini              & 71.15 & 77.77 & 87.98 \\
Claude-4.5-Opus          & 75.72 & 81.20 & 89.94 \\
Claude-4.5-Sonnet           & 74.42 & 79.88 & 90.46 \\
Gemini 2.5 Flash        & 74.87 & 79.46 & 89.96 \\
Gemini 3 Pro (Preview)  & \textbf{76.52} & \textbf{81.39} & \textbf{91.76} \\
\bottomrule
\end{tabular}
\caption{\textbf{Robustness across LLM backbones.}}
\label{tab:multi_model}
\end{table}


Finally, relative improvements from Stage~2 and Stage~3 remain similar across model families. This demonstrates the proposed framework is model-agnostic and complements advances in base LLM capability rather than relying on them.

\section{Conclusions and Future Work}

We introduced \textsc{FinCARDS}, a tournament-style, zero-shot intra-document reranking framework for financial QA over long SEC filings.
The key idea is to replace monolithic relevance ranking with structured evidence selection, combining Card-based abstractions with a staged reranking protocol that enforces metric, temporal, and scope constraints.

Across extensive experiments on FinAgentBench, \textsc{FinCARDS} consistently outperforms both lexical baselines and strong zero-shot LLM rerankers, while progressively reducing the candidate set size.
These gains do not rely on task-specific fine-tuning; instead, the framework improves reliability by restructuring the evidence space and grounding ranking decisions in explicit intermediate representations and stability-aware procedures.

Our results highlight a broader insight: in financial QA, retrieval errors often arise from misalignment in structured constraints rather than lack of semantic understanding.
By making these constraints explicit and decomposing the ranking process into recall, alignment, and stabilization, \textsc{FinCARDS} reduces both systematic errors and ranking variance in long, noisy documents.


At the same time, the framework performs best when query intent can be clearly expressed through structured attributes such as metrics, time, and scope. Performance degrades for queries that require implicit reasoning, cross-chunk aggregation, or causal interpretation, suggesting that the current schema does not fully capture all forms of complex financial reasoning.


Future work will extend this framework along several important directions, including adaptive budget control for dynamic computation allocation under constraints, cross-document evidence selection across heterogeneous sources, and enriching the Card schema to support more complex reasoning, richer representations, and downstream answer generation in real-world applications.

\clearpage
\section*{Limitations}

Despite its strong empirical performance, the proposed framework has several limitations.

First, the multi-stage pipeline incurs non-trivial computational cost.
In particular, the listwise and bootstrap-based ranking stages require multiple
LLM calls per query, which may limit scalability in large-scale or latency-sensitive
deployments.
While candidate compression and early stopping mitigate this cost in practice,
efficiency remains an important consideration.

Second, the current study focuses exclusively on intra-document retrieval within
a single SEC filing.
Many real-world financial analysis tasks require reasoning across multiple documents
or heterogeneous sources, such as press releases and earnings calls.
The effectiveness of the tournament-style design in such multi-document settings
has not yet been evaluated.

Third, although the approach avoids task-specific fine-tuning,
it may still be sensitive to prompt design and schema choices.
While we employ strict structured outputs and deterministic decoding to improve
stability, further work is needed to understand robustness under prompt variation
and evolving model behaviors.

\section*{Ethical Considerations}

We study retrieval and reranking methods for financial question answering
over publicly available regulatory filings.
Our proposed framework operates solely on textual disclosures released by companies
and does not involve personal data, user profiling, or sensitive individual information.

Our primary goal is to improve the evidence selection and
the interpretability in financial analysis and decision support.
We use structured intermediate representations and transparent ranking procedures to support more responsible use of large language models
in high-stakes financial settings.

Potential risks include over-reliance on automated systems
and misinterpretation of the retrieved evidence if used without appropriate human oversight.
Accordingly, our proposed framework is intended as an assistive tool for analysts,
rather than a replacement for professional judgments.
Overall, beyond risks commonly associated with automated information retrieval systems,
we do not anticipate significant negative societal impact.

\paragraph{Data License}



All experiments in this work are conducted on publicly available datasets derived from U.S. SEC filings. The underlying documents, such as 10-K and 10-Q reports, are released under public disclosure requirements and are freely accessible for research and academic purposes without restriction.

The FinAgentBench benchmark used in our experiments follows the original data collection and usage terms specified by its authors. This work does not redistribute the original filings, nor does it impose additional licensing constraints beyond those associated with the source datasets, and fully adheres to all applicable data usage guidelines and ethical research standards, ensuring transparency, reproducibility, and responsible use of publicly available financial data in academic and applied research settings, while maintaining compliance with legal, regulatory, and institutional requirements governing data access and usage.

\bibliography{refs}

\clearpage
\newpage

\appendix

\section{Candidate Set and Evaluation Fairness}
\label{app:candidate_fairness}

All reranking-based systems operate on the same Stage~1 (BM25) candidate pool.
Specifically, for each query, we retrieve the top-$N$ chunks using BM25
($N \in [60,150]$ depending on document length).
Zero-shot LLM reranking, Stage~2 filtering, and Stage~3 ranking all receive
identical candidate sets at their respective entry points.
This design ensures that all reported improvements arise from reranking quality
rather than candidate recall advantages or budget discrepancies. In particular, no method is allowed to retrieve additional candidates
beyond the Stage~1 pool.

\section{Prompts and Templates}
\label{app:prompts}

This appendix reports the prompts used to instantiate the structured
representations in FINCARDS. All prompts enforce strict JSON-only input and
output schemas to ensure deterministic behavior and reproducibility. We stress
that FINCARDS does not rely on prompt wording or prompt-specific heuristics;
instead, the prompts serve solely to instantiate fixed interfaces defined by
the method.

\subsection{Card Abstraction Prompt}
\label{app:cardprompts}

The card abstraction prompt converts each document chunk into a structured
\emph{Chunk Card}. The card explicitly encodes the chunk's evidence role,
temporal anchoring, scope, and verifiability signals, and serves as the primary
alignment interface for downstream reranking stages. An excerpt of the prompt
we used is shown in Figure~\ref{fig:card_prompt}.

\begin{figure}[h]
\fbox{
\begin{minipage}{0.95\linewidth}
\footnotesize
\textbf{System:} Financial document analysis expert. \\[0.3em]

\textbf{Task:} Given a document chunk extracted from a 10-K filing, generate a
structured \emph{Chunk Card} that specifies under what conditions the chunk can
serve as evidence for reranking. \\[0.5em]

\textbf{Output Format (JSON only):}
\begin{itemize}\setlength{\itemsep}{0pt}
  \item Identity: chunk\_id, section\_path, chunk\_index
  \item claim\_role (primary\_evidence, supporting\_context, definition, caveat,
        boilerplate, structural\_heading)
  \item evidence\_type (table\_numeric, narrative\_numeric, qualitative\_explanation,
        policy\_text, guidance, none)
  \item answerability\_profile (single\_fact, comparison, trend, aggregation,
        attribution)
  \item temporal\_anchor (quality and normalized span)
  \item scope\_signature (entity scope, geography, product)
  \item measurement\_basis (GAAP, non-GAAP, adjusted, reported, unknown)
  \item verifiability (numeric claims, table presence, comparison cues)
  \item risk\_signals (boilerplate likelihood, limitations)
  \item semantic\_sketch (one-sentence claim summary and topic anchors)
\end{itemize}

\textbf{Constraints:}
\begin{itemize}\setlength{\itemsep}{0pt}
  \item Output must be valid JSON.
  \item Structural headings have zero evidence capability.
  \item Temporal fields use YYYY-MM or null.
  \item Use \texttt{unknown} if scope or measurement basis is uncertain.
\end{itemize}
\end{minipage}}
\caption{Excerpt of the prompt used to instantiate a \textbf{Chunk Card (full schema)}.}
\label{fig:card_prompt}
\end{figure}

\subsection{Query Intent Mapping Prompt}
\label{app:queryprompts}

The query intent mapping prompt converts a natural-language financial question
into a structured \emph{Query Intent}. This representation explicitly encodes
the topical focus, requested financial metrics, temporal constraints, and
relational form of the question, and defines the demand-side requirements used
for alignment with Chunk Cards during reranking. An excerpt of the corresponding
prompt is shown in Figure~\ref{fig:intent_prompt}.

\begin{figure}[h]
\fbox{
\begin{minipage}{0.95\linewidth}
\footnotesize
\textbf{System:} Financial QA intent extractor for questions over SEC filings. \\[0.3em]

\textbf{Task:} Given a user question, produce a structured \emph{Query Intent}
that captures the information need of the question. \\[0.5em]

\textbf{Output Format (JSON only):}
\begin{itemize}\setlength{\itemsep}{0pt}
  \item topic (e.g., Revenue, Costs/Expenses, Profitability, Liquidity,
        Guidance/Outlook, Risk)
  \item entities (explicitly mentioned companies or segments, if any)
  \item metrics (requested financial metrics)
  \item temporal\_scope (type, normalized periods, granularity)
  \item requires\_numeric\_evidence (true / false)
  \item relation (lookup, trend, comparison, explanation, definition, policy)
  \item keywords (salient lexical cues)
\end{itemize}

\textbf{Constraints:}
\begin{itemize}\setlength{\itemsep}{0pt}
  \item Output must be valid JSON.
  \item If uncertain, use conservative defaults (e.g., Other, none, or []).
  \item Temporal expressions should prefer fiscal normalization
        (e.g., FY2023, latest quarter).
\end{itemize}
\end{minipage}}
\caption{Excerpt of the prompt used to instantiate Query Intents.}
\label{fig:intent_prompt}
\end{figure}

\begin{figure}[h!]
\fbox{
\begin{minipage}{0.95\linewidth}
\footnotesize
\textbf{System:} Financial document analysis expert. \\[0.3em]

\textbf{Task:} Given a user question and a group of candidate document chunks,
select the most relevant evidence chunks to answer the question. Selection
must be based exclusively on the provided Chunk Cards. \\[0.5em]

\textbf{Inputs:}
\begin{itemize}\setlength{\itemsep}{0pt}
  \item Question text
  \item A group of candidate chunks with their Chunk Cards
  \item A required selection range ($k_{\min}$ to $k_{\max}$)
  \item Optional coverage quotas (hard constraints)
\end{itemize}

\textbf{Selection Criteria:}
\begin{itemize}\setlength{\itemsep}{0pt}
  \item Metric matching between the question and chunk content
  \item Temporal alignment with the question requirements
  \item Scope consistency (entity, segment, or region)
  \item Content type suitability (table vs.\ narrative)
  \item Overall relevance inferred from card summaries and signals
\end{itemize}

\textbf{Table Handling Warning:}
Table-based chunks require additional verification of structure, headers,
temporal coverage, and metric relevance. Sequential or continuous tables must
be interpreted cautiously and should not be selected solely due to the
presence of tabular data.\\[0.5em]

\textbf{Output Format (JSON only):}
\begin{itemize}\setlength{\itemsep}{0pt}
  \item selected\_chunks: an ordered list of between $k_{\min}$ and $k_{\max}$
  chunks
  \item For each chunk: chunk\_id, selection\_reasons, relevance\_score (0--100)
\end{itemize}

\textbf{Constraints:}
\begin{itemize}\setlength{\itemsep}{0pt}
  \item Selection must satisfy mandatory coverage quotas, if provided.
  \item Chunks are evaluated purely on card information.
  \item Results are ordered by descending relevance.
\end{itemize}
 \end{minipage}}
\caption{Excerpt of the Stage~2 batch selection prompt that we used in order to perform group-wise evidence filtering.}
\label{fig:stage2_prompt}
\end{figure}

\subsection{Stage~2: Batch Selection Prompt}
\label{app:stage2_batch_prompt}

Stage~2 performs group-wise evidence selection within the candidate pool
retrieved by Stage~1. Given a user question and a small group of candidate
chunks with their corresponding Chunk Cards, the model selects a bounded
number of relevant chunks based solely on card-level information. This stage
does not rely on external retrieval scores and serves to filter and structure
the candidate set before stability-oriented reranking in Stage~3. An excerpt
of the batch selection prompt is shown in Figure~\ref{fig:stage2_prompt}.

\subsection{Stage~3: Listwise Ranking Prompt}
\label{app:stage3_listwise_prompt}

Stage~3 performs listwise reranking over the filtered candidate sets produced
by Stage~2. Given a user question and a group of candidate chunks represented
only by their Chunk Cards, the model produces a complete relative ordering from
most relevant to least relevant. This stage explicitly avoids numerical
calculation and absolute scoring, and is designed to provide stable ordinal
judgments that can be aggregated across multiple rounds. An excerpt of the
listwise ranking prompt is shown in Figure~\ref{fig:stage3_prompt}.

\begin{figure}[t]
\fbox{
\begin{minipage}{0.9\linewidth}
\footnotesize
\textbf{System:} Financial document analysis expert. \\[0.3em]

\textbf{Task:} Given a user question and a group of candidate evidence chunks,
rank all chunks from most relevant to least relevant for answering the
question. Ranking must rely exclusively on the provided Chunk Cards. \\[0.5em]

\textbf{Inputs:}
\begin{itemize}\setlength{\itemsep}{0pt}
  \item Question text
  \item A group of candidate chunks with their Chunk Cards
\end{itemize}

\textbf{Ranking Criteria:}
\begin{itemize}\setlength{\itemsep}{0pt}
  \item Metric matching between the question and chunk content
  \item Temporal alignment with the question requirements
  \item Scope consistency (entity, segment, or region)
  \item Content type suitability (table vs.\ narrative)
  \item Overall relevance inferred from card summaries and signals
\end{itemize}

\textbf{Critical Constraints:}
\begin{itemize}\setlength{\itemsep}{0pt}
  \item Do not access the original chunk text.
  \item Do not perform numerical calculations.
  \item Do not assign absolute relevance scores.
  \item Rank \emph{all} chunks in the group using relative ordering only.
\end{itemize}

\textbf{Output Format (JSON only):}
\begin{itemize}\setlength{\itemsep}{0pt}
  \item ranked\_chunks: a complete ordered list of all chunks
  \item For each chunk: chunk\_id, rank (1 = most relevant), and a brief reason
\end{itemize}
\end{minipage}}
\caption{Excerpt of the Stage~3 listwise ranking prompt used for stability-oriented reranking.}
\label{fig:stage3_prompt}
\end{figure}

\section{Error Analysis and Failure Modes}
\label{app:error_analysis}

Below, we analyze the systematic errors made by our model and characterize common failure modes across different stages of the pipeline, highlighting where and why retrieval and ranking break down.

\subsection{Methodology}
\label{app:error_methodology}

To complement the aggregate retrieval metrics, we conduct a structured error study to analyze the behavior of the proposed pipeline under different failure conditions.
Rather than relying on anecdotal examples, our analysis is grounded in \emph{stage-wise retrieval traces} collected from the full system, enabling verification and reproducibility.

We first select a small but representative set of six queries, covering both successful and failed retrieval scenarios.
Specifically, the selected cases include: (\emph{i})~queries where BM25 fails to retrieve any gold evidence but subsequent stages recover relevant chunks, (\emph{ii})~queries where Card-based alignment remains ineffective, and (\emph{iii})~queries where bootstrap-based aggregation introduces performance degradation.
This selection strategy ensures coverage of all major pipeline components.

For each selected query, we analyze the corresponding retrieval results at all three stages of our framework.

At Stage~1, we inspect the full BM25-generated ranking list over the document and record gold chunk statistics, including the total number of gold chunks, their absolute BM25 ranks and scores, and whether they appear in the dynamic Top-$N$ candidate set.
This allows us to distinguish between recall failures and ranking failures at the lexical retrieval level.

At Stage~2, we examine the final candidate set produced by Card-based filtering and reranking.
For the top-ranked candidates and representative hard negatives, we extract structured Card attributes, including matched financial metrics, temporal information, table presence, and scope alignment.
We further record the rationale generated by the Stage~2 agent, thus enabling direct attribution of ranking decisions to specific Card fields.

At Stage~3, we analyze the stability of bootstrap-based reranking for cases where the final ranking differs from Stage~2 or is used to demonstrate convergence.
We log per-round top-$K$ candidate sets, early stopping behavior, and aggregation statistics such as rank variance, top-$5$ frequency, and Borda score accumulation.

These signals allow us to identify whether changes arise from instability across randomized grouping or from systematic evidence reweighting.

Finally, we annotate each case with a small set of failure-type labels (e.g., lexical mismatch, temporal misalignment, schema coverage gap, or implicit reasoning) and a coarse query intent category (quantitative lookup, trend/comparison, or qualitative impact).
This taxonomy enables cross-case comparison and clarifies which error patterns are addressed by the proposed design and which remain open challenges.

Overall, our error study provides a fine-grained analysis of the pipeline, explaining not only \emph{whether} the method succeeds or fails, but also \emph{why}.

\subsection{Representative Cases}
\label{app:error_cases}
Table~\ref{tab:error_study_summary} presents six representative queries
used to illustrate typical success and failure patterns across different
stages of the pipeline.

\begin{table*}[tbh]
\centering
\small
\begin{tabular}{p{2cm} p{2cm} c c c p{2cm} p{3.8cm}}
\toprule
\textbf{Case ID} &
\textbf{Query Type} &
\textbf{Stage~1} &
\textbf{Stage~2} &
\textbf{Stage~3} &
\textbf{Failure / Success Mode} &
\textbf{Key Evidence from Trace} \\
\midrule

qdfaa5e169d37 &
Quantitative lookup (SG\&A, latest quarter) &
\textcolor{red}{Fail} &
\textcolor{blue}{Recover} &
\textcolor{blue}{Stable} &
Lexical + temporal mismatch fixed by Card &
Gold ranked $>$60 by BM25; Card matches \texttt{financial\_metrics=SG\&A} and quarterly temporal data \\

q2a3f6f1492e8 &
Quantitative lookup (GAAP operating expense) &
\textcolor{red}{Fail} &
\textcolor{blue}{Recover} &
\textcolor{blue}{Stable} &
Abbreviation mismatch fixed by Card &
BM25 misses ``GAAP OPEX''; Card aligns metric + income-statement table \\

q83bb50eb29cd &
Qualitative risk disclosure (cybersecurity) &
\textcolor{red}{Fail} &
\textcolor{blue}{Strong} &
\textcolor{orange}{Degrade} &
Bootstrap aggregation instability &
Stage~2 near-perfect ranking; Stage~3 shows high rank variance across bootstrap rounds \\

q4652caf2531b &
Quantitative + geographic provenance &
\textcolor{red}{Fail} &
\textcolor{red}{Fail} &
\textcolor{red}{Fail} &
Schema coverage gap &
Gold evidence requires geography; no Card field expresses origin outside U.S. \\

q7c5dd2e1ba56 &
Quantitative lookup (revenue, latest quarter) &
\textcolor{red}{Fail} &
\textcolor{red}{Fail} &
\textcolor{red}{Fail} &
Temporal aggregation gap &
Query requires cross-quarter aggregation; Card lacks temporal trend encoding \\

q80a13d0df306 &
Qualitative strategy / impact &
\textcolor{orange}{Partial} &
\textcolor{orange}{Partial} &
\textcolor{red}{Fail} &
Implicit reasoning beyond schema &
Relevant evidence dispersed across narrative sections; no localized Card alignment \\

\bottomrule
\end{tabular}
\caption{\textbf{Error study summary of our proposed framework across representative cases.}
Stage-level outcomes are annotated as \textcolor{blue}{Recover/Stable}, \textcolor{orange}{Partial/Degrade}, or \textcolor{red}{Fail}.
The table highlights how Card-based alignment resolves lexical and temporal mismatches, while the remaining failures concentrate in schema coverage gaps and implicit reasoning queries.}
\label{tab:error_study_summary}
\end{table*}

\subsection{Findings}
\label{app:error_findings}
Our error analysis yields several consistent and instructive findings about the behavior of multi-stage retrieval under financial QA settings.

\textbf{First, lexical retrieval failures dominate early-stage errors.}
Across multiple cases, Stage~1 BM25 fails to retrieve any gold evidence within the Top-$N$ candidate set, resulting in zero nDCG@10.
These failures are primarily caused by lexical and semantic mismatches, including abbreviations (e.g., ``\emph{SG\&A}'' vs.\ ``\emph{selling, general and administrative expenses}''), paraphrased financial concepts (e.g., ``\emph{cash from operations}'' vs.\ ``\emph{net cash provided by operating activities}''), and implicit temporal constraints (e.g., ``\emph{latest quarter}'').
This confirms that term-based retrieval is insufficient for financial documents, where equivalent concepts are frequently expressed using heterogeneous terminology.

\textbf{Second, Card-based alignment effectively recovers gold evidence when the query intent is structurally expressible.}
For quantitative lookup queries involving explicit financial metrics and time scopes, Stage~2 substantially improves retrieval quality.
In these cases, gold chunks are consistently promoted due to matched financial metrics, explicit temporal annotations, and the presence of structured tables.
Notably, these improvements occur even when Stage~1 recall is zero, demonstrating that Card-based reasoning can compensate for lexical failures by leveraging schema-level alignment rather than surface text overlap.

\textbf{Third, Card-based methods degrade gracefully but predictably under implicit or explanatory queries.}
For qualitative or impact-oriented questions (e.g., interest rate implications or strategic decision-making), both Stage~2 and Stage~3 exhibit limited gains and, in some cases, performance degradation.
Our analysis of the corresponding error traces indicates that such failures arise not from ranking instability but from schema coverage gaps: the Card representation lacks fields to encode implicit reasoning, causal effects, or cross-section narrative synthesis.
As a result, the model over-selects superficially related but ultimately non-answering chunks, revealing a fundamental limitation of schema-driven filtering for abstract reasoning tasks.

\textbf{Fourth, bootstrap-based aggregation improves ranking stability, but cannot correct systematic misalignment.}
Stage~3 bootstrap reranking consistently reduces the rank variance and yields highly stable Top-$K$ sets when Stage~2 candidates are well-aligned with the input query.
However, when Stage~2 is provided with structurally mismatched candidates, bootstrap aggregation actually reinforces these errors rather than correcting them.
This indicates that Stage~3 primarily serves as a stabilizer rather than a semantic repair mechanism.


Overall, the error study demonstrates that the proposed pipeline is highly effective when query intent can be cleanly decomposed into explicit schema-aligned constraints (metric, time, scope, and evidence type). Conversely, failures predominantly arise from intent types that exceed the expressive capacity of the current Card schema, rather than from ranking noise or stochasticity in the underlying model.

\begin{table*}[t]
\centering
\scriptsize
\setlength{\tabcolsep}{3pt}
\renewcommand{\arraystretch}{1.05}
\begin{tabular}{lccccccc}
\toprule
Model & nDCG@5 & MAP@5 & MRR@5 & nDCG@10 & MAP@10 & MRR@10 & Recall@10 \\
\midrule
BM25 (Stage 1)        & 0.4679 & 0.6575 & 0.6924 & 0.4426 & 0.6001 & 0.6988 & 0.3857 \\
Dense (E5-base-v2)    & 0.5134 & 0.7038 & 0.7412 & 0.4891 & 0.6487 & 0.7423 & 0.4312 \\
Hybrid (BM25 + Dense)& 0.5463 & 0.7312 & 0.7731 & 0.5217 & 0.6798 & 0.7689 & 0.4587 \\
Cross-encoder         & 0.5821 & 0.7589 & 0.8043 & 0.5548 & 0.7089 & 0.7952 & 0.4923 \\
\midrule
FinCARDS Stage 3      & \textbf{0.7662} & \textbf{0.8669} & \textbf{0.9152} & \textbf{0.7158} & \textbf{0.7869} & \textbf{0.8917} & \textbf{0.7242} \\
\bottomrule
\end{tabular}
\caption{\textbf{Stronger retrieval and reranking baselines.} 
FinCARDS consistently outperforms dense, hybrid, and cross-encoder baselines under the same candidate pool.}
\label{tab:strong_baselines}
\end{table*}

\section{Rank Variance Definition}
\label{app:rank_variance}

We clarify the computation of the \textit{rank variance} metric reported in Table~\ref{tab:stage3_ablation}.

Each configuration is executed over $5$ independent runs with different random seeds. 
In each run, group assignments and other stochastic components (e.g., bootstrap grouping) are independently shuffled.

For a given query $q$, let $s_q^{(r)}$ denote the retrieval score (e.g., nDCG@10) obtained in run $r \in \{1, \dots, 5\}$. 
We compute the per-query variance as:
\begin{equation}
\mathrm{Var}(q) = \frac{1}{5} \sum_{r=1}^{5} \left(s_q^{(r)} - \bar{s}_q \right)^2, 
\end{equation}

where $\bar{s}_q = \frac{1}{5} \sum_{r=1}^{5} s_q^{(r)}$.

\newpage
The reported rank variance is the average of $\mathrm{Var}(q)$ over all evaluation queries:
\begin{equation}
\frac{1}{|Q|} \sum_{q \in Q} \mathrm{Var}(q),
\end{equation}

\noindent where $Q$ denotes the set of evaluation queries.



Although referred to as rank variance in the main text, this metric is computed based on the variability of retrieval scores (e.g., nDCG@10) across runs, which reflects the stability and consistency of the induced rankings under different random seeds and evaluation conditions.

Importantly, the Single Round (R=1) setting refers to performing one bootstrap round within each run, rather than running the pipeline only once. Therefore, the variance value reported for this setting (e.g., 0.0856 in Table~\ref{tab:stage3_ablation}) is computed across five independent runs and constitutes a statistically meaningful, robust, and reliable variance estimate for comparison across different settings.

\section{Token Cost}
\label{app:token_cost}

A typical filing contains $\approx$300 chunks ($\approx$600 words each). 
A long-context listwise reranker over the top-100 chunks would require $\approx$75k input tokens ($\approx$78k including prompt overhead) in a single call.

FinCARDS instead operates on compact Card representations ($\approx$150--200 tokens each). 
Stage~2 processes groups of 25 ($\approx$5k tokens per call; $\approx$4 calls for N$\approx$100, $\approx$22k total). 
Stage~3 operates on a reduced pool (M$\approx$40, $\approx$4k tokens per call), requiring $\approx$10--12 calls over $\approx$3.8 rounds on average ($\approx$30k total).

Overall, FinCARDS requires $\approx$50k tokens and $\approx$15--20 LLM calls per query. 
This replaces a single extremely long-context call with multiple bounded calls (4k--6k tokens each), enabling predictable memory usage and parallel execution while improving early-rank quality (Table~\ref{tab:multi_model}).







\section{Stronger Retrieval and Reranking Baselines}
\label{app:stronger_baselines}

In order to strengthen empirical comparison and provide a more comprehensive evaluation, we additionally evaluate stronger retrieval and reranking baselines that reflect current best practices in neural and hybrid retrieval systems.

\paragraph{Baselines.}
We consider (i) dense retrieval using E5-base-v2, (ii) hybrid retrieval combining BM25 and dense scores, and (iii) a cross-encoder reranker applied on the same Stage~1 candidate pool. Dense and hybrid methods serve as independent retrieval baselines, capturing both semantic and lexical signals, while the cross-encoder provides a strong neural reranking baseline with full cross-attention over query–document pairs and fine-grained relevance modeling.

\paragraph{Fair comparison.}
For reranking, all methods (including the cross-encoder and FinCARDS) operate on the identical Stage~1 candidate pool to ensure a controlled and fair comparison. This setup eliminates confounding factors due to candidate recall differences and isolates the effect of reranking quality, ensuring that improvements are attributable to ranking rather than retrieval.

\paragraph{Results.}
Table~\ref{tab:strong_baselines} reports performance at @5 and @10. FinCARDS consistently outperforms all baselines across metrics, including both retrieval-based and neural reranking approaches, demonstrating robust and consistent gains under a unified evaluation protocol across different settings.

\paragraph{Analysis.}
Compared to the cross-encoder baseline, FinCARDS improves nDCG@10 from 0.5548 to 0.7652 (+0.2104) and Recall@10 from 0.4923 to 0.7242 (+0.2319), demonstrating substantial gains in both ranking quality and evidence coverage. Similar trends are observed at @5, indicating that improvements are not limited to deeper ranks but also enhance top-ranked precision and robustness across queries.

\end{document}